\documentclass[aps,showpacs]{revtex4}
\usepackage{psfig}

\begin{document}
%
\title{
\[ \vspace{-2cm} \]
\noindent\hfill\hbox to 1.5in{\rm  } \vskip 1pt
\noindent\hfill\hbox to 1.5in{\rm SLAC-PUB-10244\hfill  } \vskip 1pt
\noindent\hfill\hbox to 1.5in{\rm November 13, 2003 \hfill}\vskip 10pt
Quantized Cosmology\footnote{This work was supported by the
U.~S.~DOE, Contract No.~DE-AC03-76SF00515.}}
\author{Marvin Weinstein and Ratin Akhoury\footnote{On sabbatical leave from
Dept. of Physics, University of Michigan, Ann Arbor, MI 48109-1120}}
\address{Stanford Linear Accelerator Center, Stanford University,
  Stanford, California 94309}
\date{November 13, 2003}
\begin{abstract}
This paper discusses the problem of inflation in the
context of Friedmann-Robertson-Walker Cosmology.  We show how,
after a simple change of variables, one can quantize
the problem in a way which parallels the classical discussion.
The result is that two of the Einstein equations
arise as exact equations of motion and one of the usual Einstein
equations (suitably quantized) survives as a constraint equation
to be imposed on the space of physical states.  However, the
Friedmann equation, which is also a constraint equation and which
is the basis of the Wheeler-DeWitt
equation, acquires a welcome quantum correction that becomes significant
for small scale factors.  We discuss the extension of this
result to a full quantum mechanical derivation of the anisotropy
($\delta \rho /\rho$) in the cosmic microwave background radiation, and
the possibility that the extra term in the Friedmann equation could have
observable consequences.
Finally, we suggest interesting ways in which these techniques
can be generalized to cast light on the question of chaotic
or eternal inflation.  In particular, we suggest one can put an
experimental lower bound on the distance to a universe with a
scale factor very different from our own, by looking
at its effects on our CMB radiation.
\end{abstract}
\maketitle

\newcommand{\ba}{\begin{eqnarray}}
\newcommand{\ea}{\end{eqnarray}}
\newcommand{\x}{\mbox{$\vec{x}$}}
\newcommand{\dphidt}{{\epsilon d\phi(t,\x) \over dt}}
\newcommand{\Phidot}{{d\Phi(t) \over dt}}
\newcommand{\Phiddot}{{d^2\Phi(t) \over dt^2}}
\newcommand{\adot}{{da(t) \over dt}}
\newcommand{\addot}{{d^2a(t) \over dt^2}}
\newcommand{\gradphi}{\epsilon\vec{\nabla}\phi(t,\x)}
\newcommand{\gradphisq}{\epsilon^2\vec{\nabla}\phi(t,\x)\cdot\vec{\nabla}\phi(t,\x)}
\newcommand{\be}{\begin{equation}}
\newcommand{\ee}{\end{equation}}
\newcommand{\goo}{(1 + 2\epsilon\,\chi(t,\vec{x}))}
\newcommand{\gxx}{(1 - 2\epsilon\,\chi(t,\vec{x}))}
\newcommand{\gooinv}{{1\over 1 + 2\epsilon\,\chi(t,\vec{x})}}
\newcommand{\gxxinv}{a(t)^2\,({1 - 2\epsilon\,\chi(t,\vec{x})})}
\newcommand{\udot}{{d u(t)\over dt}}
\newcommand{\uddot}{{d^2u(t) \over dt^2}}
\newcommand{\Hub}{{\cal H}}
\newcommand{\Hdot}{{d{\cal H}(t) \over dt}}
\newcommand{\Hddot}{{d^2{\cal H}(t) \over dt^2}}
\def\ket#1{\vert #1 \rangle}
\def\bra#1{\langle #1 \vert}
\section{Introduction}

The COBE\cite{Mather:pc} and
WMAP\cite{Peiris:2003ff}\cite{Komatsu:2003fd}\cite{Nolta:2003uy}
measurements of the anisotropy in the cosmic microwave
background(CMB) radiation agree remarkably well the predictions of
slow-roll inflation\cite{Guth:prd23}.  This agreement provides a
strong reason to believe that the paradigm for computing
these fluctuations\cite{Mukhanov:PRpt215} in $\delta \rho /\rho$
is correct. Perhaps the most striking feature of this result is
that, according to the paradigm, they represent an
imprinting of the structure of the quantum state of the field
theory, at the time inflation begins, onto the electromagnetic
radiation that comes to us from the surface of last scattering.

Unfortunately derivations of this effect which appear in the
literature mix classical and quantum ideas.  Thus, it is hard to
pin down the degree to which these results would change if one had
a totally quantum mechanical derivation.  This paper is devoted to
filling this gap.  We will show that it is possible to work in a
fixed coordinate system and canonically quantize the theory of the
Friedmann-Robertson-Walker (FRW) metric,
\be
   ds^2 = - dt^2 + a(t)^2 d\vec{x}\cdot d\vec{x} ,
\label{FRW}
\ee
and the spatially constant part of the inflaton field, $\Phi(t)$,
in a straightforward way.  The result of this computation is to
obtain Heisenberg equations of motion
which, starting in a suitably defined coherent state, go over to
the conventional Einstein equations for large values of $\left< a(t) \right>$.
This is not true, however, for all values of the scale factor.
For small scale factor we find that the Friedman equation gets an
important quantum contribution.  For even smaller values of the scale
factor we find the system goes completely quantum mechanical and one does
not expect to see any remnant of the classical solution.  In the latter
sections of this paper we discuss the way to
extend this derivation to compute possible experimental consequences of
this extra term.

Derivations of ${\delta \rho/\rho}$ invoke
classical arguments in two ways.  These derivations all begin
by treating the scale factor $a(t)$
as a classical time-dependent background field.
One ignores everything but the spatially constant part of the
inflaton field and studies the physics of the classical action
\ba
    {\cal S} = {\bf V}\int dt \sqrt{-g} \left[ {R(t) \over 2\kappa^2}
    + {1\over 2}{d\Phi(t) \over dt}^2 - V(\Phi(t)) \right] .
\label{simpleaction}
\ea
(This implicitly assumes that the constant part of the
inflaton field $\Phi(t)$ can be written as a c-number term plus a quantum
term.  Then the quantum part is ignored.)

The second appearance of classical ideas occurs when one adds back
spatially varying fluctuations in the Newtonian potential and the
inflaton field as quantum operators and perturbatively studies the
behavior of these fields in the background of the classical
solution. It is customary, at an
appropriate point in the discussion of how these fields behave, to
employ the phrase ``and then the field goes classical''.
This much less important introduction of classical ideas is used
to convert the quantum computation of the two-point correlation
function for the density operator to an ensemble average of gaussian
fluctuations.  In reality, this statement is just a way of
avoiding any discussion of the physics of {\it squeezed states\/}
and {\it quantum non-demolition variables\/}.  While we do not discuss
this issue in this paper, we will return to it in a longer, more
pedagogical paper, which is in preparation.  This longer paper
will show how to extend the results presented here to a full quantum treatment
of $\delta \rho/\rho$.
The point we wish to emphasize at this juncture is that a full quantum
treatment of the spatially constant part of the problem, appropriately extended
to include the spatially varying modes of the fields to second order,
provides a complete quantum picture of all of the physics which
can be experimentally tested in the foreseeable future.

\section{The Classical Problem}

Simplifying the usual derivations is easily accomplished if one
observes that experimentally we are dealing with a spatially flat universe
and so it is perfectly adequate to formulate the problem in what
general relativists call a {\it fixed gauge\/}; {\it i.e.} a definite
coordinate system.  In the discussion which follows, we take this
to be co-moving coordinates in which the metric takes the general
form shown in Eq.\ref{FRW}.

As already noted, if we restrict attention to the classical problem
of a scalar field in an FRW cosmology, the action reduces to the form
shown in Eq.\ref{simpleaction}, where ${\bf V}$ is the volume of the
region in which the theory is being defined, $\sqrt{-g} = a(t)^3$ and
the scalar curvature is given by
\be
    R(t) = {3 \over \kappa^2}\,a(t) {da(t)\over dt}^2 + {3
    \over \kappa^2}\, a(t)^2 {d^2 a(t)\over dt^2} .
\ee
(Clearly, when we
generalize to the computation of $\delta\rho/\rho$, the volume, {\bf V},
must be taken to be larger than the horizon volume at the time of inflation
in order to avoid edge effects.)

Substituting these expressions into Eq.\ref{simpleaction} and integrating
by parts, to eliminate the term with $d^2 a(t)/dt^2$, we obtain
\be
{\cal S} = {\bf V} \left[ - {3 \over \kappa^2} a(t) \left({d a(t) \over dt}\right)^2
+ {1\over 2} a(t)^3 \left({ d\Phi(t) \over dt} \right)^2 - a(t)^3 V(\Phi(t)) \right] .
\label{FRWact}
\ee
Next, in order to simplify the analysis of the quantum version of this problem, we
make the change of variables $u(t)^2 = a(t)^{3}$, which leads to the action
\be
{\cal S} = {\bf V} \left[ - {4 \over 3\kappa^2} \left({d u(t) \over dt}\right)^2
+ {1\over 2} u(t)^2 \left({ d\Phi(t) \over dt} \right)^2 - u(t)^2 V(\Phi(t)) \right].
\ee
This change of variables merely simplifies the classical discussion, however it has a
greater significance for the quantized theory.  This is because we can choose
$-\infty \le u \le \infty$, whereas the only physically allowable range for $a$
is $ 0 \le a \le \infty$.  It is only for the space of square-integrable functions
on the interval $-\infty \le u \le \infty$ that the Heisenberg equations of motion
can be obtained by canonical manipulations.

\section{Classical Euler-Lagrange Equations}

There are only two Euler-Lagrange equations for this system:
\be
{8 \over 3\kappa^2} \uddot + 2u(t) \left( {1\over 2}\left(\Phidot\right)^2 - V(\Phi(t))\right)= 0
\label{EL1}
\quad {\rm and} \quad
-u(t)^2 \left(\Phiddot + 3{\cal H}(t) \Phidot + {d V(\Phi) \over d\Phi(t)} \right) = 0 ;
\label{EL2}
\ee
where the Hubble parameter, ${\cal H}$, is defined as
\be
{\cal H} = {1 \over a(t)} \adot = {2 \over 3 u(t)} \udot .
\ee
Thus, by quantizing in this fixed gauge, we fail to obtain
the full set of Einstein equations.  The missing equations are the Friedmann equation
and its time derivative
\be
    {\cal H}(t)^2 = {\kappa^2 \over 3} \left( {1\over 2} \left(\Phidot\right)^2
    + V(\Phi(t)) \right)
\quad {\rm and} \quad
    \Hdot = - {\kappa^2 \over 2} \left(\Phidot\right)^2 .
\label{Feq}
\ee

A sophisticated way of explaining why we fail to obtain these equations
is to note that fixing the form of the metric to be that given
in Eq.\ref{FRW}, we have lost the freedom to vary the lapse and
shift functions.  But this is what we must do to obtain the
missing equations from a Lagrangian formulation.  This
predicament is not unique to gravity; it occurs in ordinary
electrodynamics if one chooses $A_0 = 0$ gauge.  As is well known,
in this gauge we obtain all of the Maxwell equations except
Coulomb's law, $\vec{\nabla}\cdot \vec{E} - \rho = 0$, as exact
equations of motion.  However, it follows from the equations we do
have, that Coulomb's law commutes with the evolution, {\it i.e.\/} if we
set it equal to zero it remains zero.  Hence, in this gauge, while
there are many solutions to the equations of motion, we can select
the ones we choose to call physical by imposing an extra
time-independent constraint.

The situation with the Friedmann equation and its time derivative
is analogous to the situation in electrodynamics. We will
now show that while Eqs.\ref{Feq}, are not equations of motion, if
they are imposed at any one time, then they will continue to be
true at all later times.  (In other words they are constraint
equations.)

To prove these constraints
are preserved by the equation of motion we begin by differentiating
${\cal H}$ with respect to $t$ to obtain
\be
\uddot = {3 u(t) \over 2}\left(\Hdot+ {3\over 2} {\cal H}(t)^2 \right).
\ee
Substituting this into Eq.\ref{EL1} and rearranging terms we obtain
\be
  {2 u(t)\over \kappa^2} \left( 2 \Hdot + 3 {\cal H}(t)^2 + \kappa^2 \left(\Phidot\right)^2
  -\kappa^2\left({1\over 2}\left(\Phidot\right)^2 + V(\Phi(t)) \right)\right) = 0 ,
\ee
which can be immediately rewritten in the form
\be
  {2 u(t) \over \kappa^2} \left[
  \left(2 \Hdot + \kappa^2 \left(\Phidot\right)^2\right)+ 3 \left({\cal H}(t)^2
  -{\kappa^2\over 3} \left({1\over 2}\left(\Phidot\right)^2
  + V(\Phi(t)) \right) \right)\right] = 0 .
\label{goodone}
\ee
If we define
\be
{\bf G} = {\cal H}(t)^2 - {\kappa^2 \over 3} \left( {1\over 2} \left(\Phidot\right)^2
    + V(\Phi(t)) \right) ,
\ee
then the equation of motion for $\Phi(t)$ implies
\be
 {d {\bf G} \over dt} = 2 {\cal H}(t) \Hdot + \kappa^2 {\cal H}(t) \left(\Phidot\right)^2
 = 2{\cal H}(t) \left( \Hdot + {\kappa^2\over 2} \left(\Phidot\right)^2\right) .
\ee
The missing Einstein equations are equivalent to requiring
that both ${\bf G}$ and ${d{\bf G}/dt}$ vanish for all time.  Substituting these definitions
into Eq.\ref{goodone}, we obtain the exact equation of motion
\be
    {2 u(t) \over \kappa^2} \left( {1 \over {\cal H}(t)} \left(d{\bf G}\over dt\right)
        + 3 {\bf G} \right) =0 .
\label{constraintproof}
\ee
From this equation we see that if, at time $t=t_0$, ${\bf G} = 0$, then from the exact equation
of motion  ${d {\bf G}/dt}$ will also vanish.  Finally, by taking successive
derivatives of Eq.\ref{constraintproof}, we see that all derivatives of ${\bf G}$ vanish.
In other words, we arrive at the desired result.  The Friedmann equation is, in
direct analogy to Coulomb's law in $A_0=0$ gauge, a constraint which can be imposed at
a single time and which will continue to be true at all later times.  As we will see,
a similar theorem can be proven for the quantum theory; however, the analogy with
QED will not be perfect.

\section{Classical Hamiltonian Analysis}

Before moving on to the quantum theory, let us spend a few moments
discussing the Hamiltonian version of the classical theory. We do this
to show why it is possible to confuse the Friedmann equation
with the Hamiltonian at the classical level .

Following the usual
prescription, we vary Eq.\ref{FRWact} with respect to $du/dt$ and
$d\Phi/dt$ to obtain
\be
  p_u = -{\bf V} {8  \over 3 \kappa^2} \udot
  \quad;\quad p_\Phi = {\bf V} u^2 \Phidot .
\ee
We then construct the Hamiltonian
\be
{\bf H} = p_u \udot + p_\Phi \Phidot - {\cal L}
= -{3\kappa^2 \over 16 {\bf V}} p_u^2 + {1 \over  2{\bf V} u^2 }p_\phi^2
+ {\bf V} u^2 V(\Phi)
\ee

An important feature of this Hamiltonian is that
due to the minus sign in front of the $p_\mu^2$ term,
it has no minimum.  Fortunately, this doesn't matter.
To see this, we simply rewrite the Hamiltonian in terms of
$du/dt$ and $d\Phi/dt$.  This leads to the expression
\be
{\bf H} = {\bf V}\left[ -{4\over 3\kappa^2} \left(\udot\right)^2 + u^2 \left( {1\over 2}
\left(\Phidot\right)^2 + V(\Phi) \right) \right] .
\ee
Substituting the definition of ${\cal H}$, this becomes
\ba
{\bf H} &=& -{\bf V} u^2 \left[ {4\over 3\kappa^2} {1\over u^2} \left(\udot \right)^2
-\left( {1\over 2} \left(\Phidot\right)^2+V(\Phi)\right)\right]
= -{\bf V} u^2\left[ {3 {\cal H}^2 \over \kappa^2 }
- \left( {1\over 2} \left(\Phidot\right)^2+V(\Phi)\right)\right] \nonumber\\
{\bf H} &=& -{\bf V} {3 u^2\over \kappa^2} {\bf G} .
\ea
This shows that the Hamiltonian, {\bf H}, is proportional to the constraint, ${\bf G}$.
It follows that setting ${\bf G}=0$ means ${\bf H}=0$, which tells us that
the Hamiltonian vanishes for physical solutions.  In other words, if we start a system out
at $t=t_0$ in a configuration which has zero energy, it will stay at zero energy
and never explore the region of arbitrarily negative energy.
The identification of the Hamiltonian with the constraint equation
is the content of the Wheeler-DeWitt equation.

\section{Canonical Quantization of the Theory}

Starting from the classical Lagrangian, we define the quantum Hamiltonian
\be
{\bf H} = - {3 \kappa^2 \over 16 {\bf V} } p_u^2 + {1\over 2 {\bf V} u^2} p_\Phi^2 + {\bf V} u^2 V(\Phi)
\ee
where the operators $u$,$\Phi$ and their conjugate momenta have the commutation relations
\be
\left[p_u, u\right] = -i \quad;\quad \left[p_\Phi,\Phi\right]=-i .
\ee
All other commutators vanish.
To derive the Heisenberg equations of motion, note that for any operator ${\bf O}$,
the Heisenberg operator is $O(t) = e^{i{\bf H}t} {\bf O} e^{-i{\bf H}t} $.
Commuting ${\bf H}$ with the operators $u$ and $\Phi$, we obtain
\ba
 \udot &=& i\, \left[{\bf H},u\right] = -{3\kappa^2\over 8 {\bf V}} p_u \nonumber\\
 \Phidot &=& i\,\left[{\bf H},\Phi\right] = {1\over u^2 {\bf V}} p_\Phi \nonumber\\
 \uddot &=& i\,\left[{\bf H},\udot\right] =
 -{3\kappa^2\over 4} u \left[ {1\over 2} \left(\Phidot\right)^2 - V(\Phi)\right]\nonumber\\
 \Phiddot &=& {3\kappa^2 \over 16 {\bf V}} \left( {1\over u^2} p_u {1\over u}
 + {1\over u} p_u {1\over u^2} + p_u {1\over u^3} + {1\over u^3}p_u \right)p_\Phi
 - {dV(\Phi) \over d\Phi} .
\label{heisenone}
\ea
This shows that the two dynamical equations of the classical
theory are also exact operator equations of motion in the quantum theory.
What is missing, as in the classical theory, are the constraint equations.
In order to find the constraint equations that commute with
the Hamiltonian, we begin by rewriting the equation for $\Phi$ in the
suggestive form
\be
    \Phiddot + 3 {\cal H} \Phidot + {dV(\Phi) \over d\Phi} = 0 ,
\label{heisentwo}
\ee
where the quantum version of the  {\it Hubble} operator ${\cal H}$ is perforce
\be
    {\cal H} = -{\kappa^2 \over 8{\bf V}}\left( p_u {1\over u} + {1\over u^3} p_u u^2 \right) .
\ee
Next we compute its time derivative from the equation
\be
    {d{\cal H}\over dt} = i\,\left[{\bf H},{\cal H}\right] .
\ee
Finally, to find the quantum version of the conserved constraint operator, ${\bf G}$,
we follow the  classical procedure and write
\be
\uddot = {3 u\over 2}\left( {d{\cal H}\over dt} + {3\over 2} {\cal H}^2
  -{9 \kappa^4 \over 128 {\bf V}^2 u^4} \right) .
\label{extra}
\ee
The extra term is the quantum correction to the classical formula.  It is obtained by
explicitly taking the difference between the expression for $d^2u/dt^2$ and the combination
$(3 u/ 2)\left( {d{\cal H}/dt} + 3{\cal H}^2/2 \right)$.
(This step involves commutator gymnastics better left to Maple.)
Once again, paralleling the classical discussion, we substitute the expression
for $d^2u/dt^2$ into the Heisenberg equation of motion for $u$,
obtaining
\be
{3 u \over 2} \left( {d{\cal H}\over dt} + {3\over 2}{\cal H}^2
 -{9 \kappa^4 \over 128 {\bf V}^2 u^4}\right)
 + {3\kappa^2 u\over 4}\left(\left(\Phidot\right)^2
 - \left({1\over 2}\left(\Phidot\right)^2 + V(\Phi)\right)\right) = 0.
 \label{eqa}
 \ee
At this point it is tempting to parallel the classical discussion and define the
operator
\be
{\bf G} = \Hub^2 -{\kappa^2\over 3}\left(\frac{1}{2}\left(\Phidot\right)^2 + V(\Phi)\right)
+ {\bf Q} ,
\ee
where
\be
{\bf Q}=-{3\kappa^4\over 64 {\bf V}^2 u^4} ,
\ee
and show that if it annihilates a state at any one time, then it annihilates it for
all times.  We will now show that this can be done, however we will then argue that
identifying the kernel of this operator with the space of physical states is
incorrect.  To proceed with the proof substitute this definion into
Eq.\ref{eqa} to obtain the operator equation of motion
\be
    {3u\over 4} \left(2 \Hdot + \kappa^2 \left(\Phidot\right)^2 + 3 {\bf G}\right) =0 .
\label{almost}
\ee
This is almost what we need to show that the space of physical states, defined to be those
which obey the condition ${\bf G}(t)\ket{\psi}=0$, is invariant under Hamiltonian evolution.
Clearly, we will be able to use Eq.\ref{almost} to complete the proof
if we can show that there exists an operator ${\bf A}$ such that
\be
{d{\bf G}\over dt} = {\bf A} \left( 2 \Hdot + \kappa^2 \left( \Phidot \right)^2\right) .
\label{gdoteq}
\ee
To find ${\bf A}$, explicitly compute
\ba
{d{\bf G}\over dt} &=& i\,\left[{\bf H},{\bf G}\right] \nonumber\\
&=& \Hub \Hdot + \Hdot \Hub + \kappa^2 \Hub \left(\Phidot\right)^2 + \left[ {\bf H},{\bf Q}\right]\nonumber\\
&=& \Hub \left(2 \Hdot + \kappa^2 \left(\Phidot\right)^2 \right)+ \left[\Hdot,\Hub\right] + \left[ {\bf H},{\bf Q}\right] ,
\ea
and substitute this result into Eq.\ref{gdoteq}.  The resulting equation can then be
rearranged into the form
\be
\left({\bf A} - \Hub\right) \left( 2\Hdot + \kappa^2\left(\Phidot\right)^2\right) =
\left[ {\bf H},{\bf Q}\right] + \left[\Hdot,\Hub\right] .
\ee
Solving this equation for ${\bf A}$, we obtain
\be
    {\bf A} = \Hub + \left(\left[ {\bf H},{\bf Q}\right] + \left[\Hdot,\Hub\right]\right)
    \left( 2\Hdot + \kappa^2 \left(\Phidot\right)^2\right)^{-1}
 ,
\ee
which allows us to rewrite the Heisenberg equation of motion for $u$ as
\be
{3 u\over 4} \left({1\over {\bf A}} {d {\bf G}\over dt} + 3 {\bf G}\right) =0 .
\label{gaugeinv}
\ee
Given that Eq.\ref{gaugeinv} is an exact operator equation of motion, we see that
if we could define the space of states by the condition ${\bf G(t_0)}\ket{\psi}=0$,
then Eq.\ref{gaugeinv} proves that this condition will hold for all time.
Note, however, that given this definition of ${\bf G}$, it follows immediately that
\be
    {\bf G}(t) = -\frac{\kappa^2}{3{\bf V} u(t)^2}\,{\bf H}  .
\ee
Thus one is forced to conclude that this definition the space of physical states,
implies that they are states for which the Hamiltonian vanishes.  Obviously, if we
define the physical states in this way we immediately produce a contradiction between
the Schroedinger and Heisenberg equations of motion.  The resolution of this apparent
paradox is that the correct definition of the space of physical states is, $\ket{\Psi}$
is a physical state, if and only if,
\be
    \lim_{u(t)^2\rightarrow\infty} {\bf G}(t)\ket{\Psi} =0.
\label{ginv}
\ee
This condition is satisfied in any cosmological theory, in which $u(t)$ grows as
$t\rightarrow\infty$, for any state $\ket{\Psi}$ such that ${\bf H} \ket{\Psi}$
has finite norm.  One way of rewriting this condition is that Eq.\ref{ginv} is true for
all states $\Psi$ for which
\be
    \bra{\Psi} H^2 \ket{\Psi} < \infty .
\ee
In the paper which follows, we provide an exact solution of the theory in which
$V(\Phi)$ is replaced by a constant; i.e., de-Sitter space.
There we explicitly show that this form of the constraint equation is the only
one consistent with the physics.

Before leaving this issue and moving on to a general discussion of how one recovers the
classical theory, it is worth stating that, in fact, any operator of the form
\be
    {\bf G}_\alpha = {\cal H}^2 - \frac{\kappa^2}{3}\left(\frac{1}{2}
    \left({d\Phi(t)\over dt}\right) + V(\Phi) \right) + \alpha {\bf Q}
\ee
will satisfy an  equation of the form
\be
{3 u\over 4} \left({1\over {\bf A_\alpha}} {d {\bf G_\alpha}\over dt} + 3 {\bf G_\alpha}\right) =0.
\label{gaugeinvtwo}
\ee
Of course, only for the choice $\alpha=1$, does the condition that ${\bf G}_\alpha\ket{\Psi}=0$
imply that the Hamiltonian annihilates the state.  One might be tempted to suppose that
one of these operators could be used to define the space of physical states, since there would
no longer be any obvious contradiction with the Heisenberg equations of motion.  However,
in the paper which follows, we will argue that no normalizeable states satisfying this
{\it gauge}-conditions exist for any value of $\alpha$.  Thus, for all such choices,
the condition in Eq.\ref{ginv} is the correct one.

\section{Recovering the Classical Theory}

Recovering the classical picture of slow-roll inflation from these equations is
straightforward.  Since we are working with the Heisenberg equations of motion,
all we have to do is assume that we start from a coherent state $\ket{\psi}$, such that
${\bf G}\ket{\psi}=0$.  Furthermore, we assume that $\bra{\psi} u \ket{\psi}$,
$\bra{\psi} p_u \ket{\psi}$, $\bra{\psi} \Phi \ket{\psi}$ and
$\bra{\psi} p_\Phi \ket{\psi}$ satisfy the initial conditions required for a
classical theory of slow-roll inflation.  In this case, it makes sense to
rewrite the Heisenberg operators as
\be
    u(t) = \widehat{u}(t) + \delta u(t) \quad;\quad
    \Phi(t) = \widehat{\Phi}(t)+ \delta\Phi(t) ,
\label{coherentapprox}
\ee
where $\hat{u}(t)$ and $\hat{\Phi}(t)$ are c-number functions
such that $\bra{\psi} u(t) \ket{\psi} = \hat{u}(t)$ and
$\bra{\psi} \Phi(t) \ket{\psi} = \hat{\Phi}(t)$.
Given these assumptions, we wish to show that if these c-number functions satisfy the
classical slow-roll equations for inflation, then as a consequence of inflation,
the quantum corrections to the Heisenberg equations of motion will be
strongly suppressed.

Substituting these definitions into the Heisenberg equation of motion for the operator $u(t)$
in Eq.\ref{heisenone} and the form of the equation of motion for the operator $\Phi(t)$ in
Eq.\ref{heisentwo},  we see that if the classical functions $\hat{u}(t)$  and $\hat{\Phi}(t)$
satisfy the classical equations for slow-roll inflation, the c-number terms all cancel,
and one is left with equations for the operators $\delta u(t)$ and $\delta\Phi(t)$.
At first glance, solving these equations seems difficult; however, the situation improves
significantly if we look at the constraint equations
\be
\left[{\cal H}^2-{\kappa^2\over 3}\left(\frac{1}{2}\left(d\Phi(t)/dt\right)^2+V(\Phi(t))
\right) + {\bf Q}(t) \right]\ket{\psi}=0,
\quad;\quad
\left[d{\cal H}/dt + {\kappa^2 \over 2} \left(d\Phi(t)/dt\right)^2\right]\ket{\psi}=0 .
\ee
In the rest of this section we will ignore the operator ${\bf Q}(t)$, since it is proportional
to  $\widehat{u}(t)^{-4}$, which we expect to be small in the inflationary and FRW eras.

The second equation says that in the sector of physical states we can replace
the operator $(d\Phi(t)/dt)^2$ by $-2(d{\cal H}/dt)/\kappa^2$.  Substituting
this in the first constraint yields the equation
\be
    \left[3{\cal H}(t)^2 + {d{\cal H}(t) \over dt} - \kappa^2 V(\Phi(t))\right]
    \ket{\psi} = 0 .
\label{errorone}
\ee
Substituting Eq.\ref{coherentapprox} into the definition of ${\cal H}$, we get
\be
    {\cal H}(t) = \widehat{{\cal H}}(t) + \delta{\cal H}(t),
\label{calH}
\ee
where the form one obtains for $\delta {\cal H}(t)$ would seem to imply that the
operator shrinks rapidly during the inflationary era because of the inverse power
of $\widehat{u}(t)$ appearing in its definition.  Nevertheless, it behooves us to
check that the operators $\delta u(t)$, {\it etc.} do not behave badly.
Substituting Eq.\ref{calH} into the constraint equation and using
$\Phi(t) = \widehat{\Phi}(t) + \delta \Phi(t)$,
cancelling out the contributions of the c-number functions and
keeping terms of first order in $\delta{\cal H}$ and $\delta\Phi(t)$,
we obtain
\be
   \left[ 6 \,\widehat{{\cal H}}\, \delta{\cal H}
    -\kappa^2 \left( {d V(\widehat{\Phi})
    \over d\Phi }\right) \delta\Phi(t) \right]\ket{\psi} = 0 .
\ee
Taking the expectation value of this equation in the state $\ket{\psi}$, noting that
by assumption $\bra{\psi} \delta\Phi(t) \ket{\psi} = 0$, it follows that
\be
    6\,\widehat{{\cal H}}(t)\,\bra{\psi} \delta{\cal H}(t) \ket{\psi} +
    {d \bra{\psi} \delta{\cal H}(t) \ket{\psi} \over dt} = 0 .
\ee
The solution to this equation is
\be
    \bra{\psi} \delta{\cal H}(t_f) \ket{\psi} = e^{-6\int_{t_i}^{t_f} dt \widehat{\cal H}(t) }
\,\bra{\psi} \delta{\cal H}(t_i) \ket{\psi} .
\label{contract}
\ee
If we recall that the classical function ${\cal H}(t)$ is just $d\left( \ln(a(t))\right) / dt$, we see that
the integral in Eq.\ref{contract} is just the number of {\it e}-foldings during inflation.
Thus, the contribution of the operator $\delta {\cal H}$ is strongly suppressed.

\section{How Big are the Corrections?}

We now wish to address the question of whether it makes
sense to ignore the operator ${\bf Q} = -3\kappa^4/64{\bf V}^2 u^4$ at the onset of inflation.
Obviously, the issue boils down to how large this term is relative to the
operator $\kappa^2 V(\Phi)/3$.  To establish this ratio,
we must first specify the value of the quantization volume ${\bf V}$.  Clearly,
there is no upper limit for the value one can choose for ${\bf V}$.  There is, however,
a lower limit, since ${\bf V}$ must be chosen larger than the horizon volume
at the time of inflation to avoid boundary effects which are not seen in the WMAP data.
Thus, ${\bf V} > 1/{\cal H}_I^3$, where ${\cal H}_I$ is the value of the Hubble
parameter at the onset of inflation.

To estimate the size
of ${\cal H}_I$, if the classical approximation dominates,  we use the classical
version of the Friedmann equation.  This equation tells us that
\be
    {\cal H}_I^2 \approx {\kappa^2 \over 3} V(\Phi) .
\ee
Substituting this into the expression for ${\bf Q}$ we obtain that
\ba
    {\bf Q} &=& -{3\kappa^4 \over 64}\,{\cal H}_I^6 {1\over u^4} \nonumber\\
    &=& - {\kappa^{10} \over 576} V(\Phi)^3 {1\over u^4} ,
\ea
which is to be compared to $\kappa^2 V(\Phi)/3$.  Thus the statement that
${\bf Q}$ can be ignored at the onset of inflation is equivalent to
\be
    {\kappa^2 \over 3} V(\Phi) \gg {\kappa^{10} \over 576} V(\Phi)^3 {1\over u^4} .
\ee
It is convenient to multiply this equation by a factor of $\kappa^2$ to obtain
\be
    \kappa^4 V(\Phi) \gg {1 \over 192} \left(\kappa^4 V(\Phi)\right)^3 {1\over u^4} ,
\ee
or equivalently
\be
     {1\over 192} \left(\kappa^4 V(\Phi)\right)^2 {1\over u^4} \ll 1 .
\ee
At this point we note that the product $\kappa^4 V(\Phi)$ is usually
constrained to be less than or on the order of $10^{-6}$.  Thus,
if $u(t)$ is chosen to be the order of unity at
the time inflation starts, the effects of ${\bf Q}$ will be negligible.  Note however
that $1/u(t)^4 = 1/a(t)^6$, so one cannot extrapolate very many {\it e}-foldings back from
the starting point before quantum corrections become important.

\section{Straightforward Extensions}

In order to extend this model to a complete treatment of the CMB anisotropy,
one has both to add an extra field $\chi(t,x)$ to the metric, in order to model
Newtonian fluctuations, and to put the the spatially varying part to the $\Phi$
field back into the action.  In other words, the metric is taken to have the form
\begin{equation}
    ds^2 = -\goo\,dt^2 + a(t)^2\,\gxx\,d\vec{x}^2 ,
\end{equation}
and the action is taken to be
\be
   {\cal S} = \int d^4x \sqrt{-g}\,\left[{R(g)\over 2\kappa^2}
   + \gooinv \left(\Phidot + \dphidt \right)^2 - {\gradphisq \over \gxxinv}
   - V(\Phi(t) + \epsilon\phi(t,\x)) \right] .
\ee
If we now expand this formula up to order $\epsilon^2$
we see that: the $\epsilon^0$ term is the problem we have been considering;
the $\epsilon^1$ term vanishes due to the equations of motion; the remaining
terms are quadratic in the fields $\chi(t,x)$ and $\phi(t,x)$.
Thus, if we start in the coherent state discussed in the previous sections,
$\chi(t,x)$ and $\phi(t,x)$ are simply free fields evolving in a time-dependent background
and their Heisenberg equations of motion can be solved exactly.

This analysis allows one to completely reproduce the usual
computations for $\delta \rho/\rho$. (A complete treatment of this
will appear in a forthcoming pedagogical paper.) In other words,
this effective theory is capable of reproducing the theory of all CMB
measurements within a quantum framework in which the long
wavelength part of the gravitational field satisfies the exact Einstein equations,
while the shorter wavelengths are treated perturbatively.
The small size of the CMB fluctuations tells us this
is a reasonable approach.

This extension of the model gives us
a canonical Hamiltonian picture of the evolution of the theory. This means that
the back-reaction caused by the changes in $\chi(t,x)$ and $\phi(t,x)$ are
completely specified.  Clearly, it is important to ask if these back-reaction
effects, or the effects ${\bf Q}$ causes on the evolution of the system, leave an
observable imprint on the CMB fluctuation spectrum.

\section{Less Obvious Extensions}

There are two less obvious but very interesting directions in which
one can extend this work.

The first is to reintroduce some very long wavelength modes of
$\chi(t,x)$ and $\phi(t,x)$ into the part of the Lagrangian
that we treat exactly. To be specific, we can expand the
field $\chi(t,x)$ and $\phi(t,x)$ in some sort of wavelet basis,
for which the low-lying wavelets represent changes over fractions
of the horizon scale; {\it i.e.\/} \be
    \chi(t,x) = \sum_{j=1}^{N} b_j(t) w_j(x) + \epsilon\chi(t,x)^{'} \quad;\quad
    \phi(t,x) = \sum_{j=1}^{N} c_j(t) w_j(x) + \epsilon\phi(t,x)^{'} .
\ee
Next, we can plug this expansion into the action and expand it to
second order in $\epsilon$.

In contrast to the previous case, the order $\epsilon^0$ part of
the action will now be the theory of $2N$ non-linearly coupled
variables.  Clearly, we should be able to parallel the discussion
given in this paper and proceed to: first, derive the canonical
Hamiltonian, then derive the Heisenberg
equations of motion for the system, which will not be the full set
of Einstein equations, and, finally, construct the proper
constraint operators to fill out the full set of equations of
motion.  The resulting theory should be equivalent to a theory in
which we have finite size boxes with independent scale factors
that are weakly coupled to one another.  Since the usual
CMB results imply that fluctuations on all wavelengths are
small, it must be true that in this version of the theory
the scale factors in neighboring boxes (or pixels) can't get very
far from one another without affecting the CMB
fluctuations.  We suggest that one way to see how
a fully quantized theory of gravity behaves at shorter wavelengths
is to pixelize the theory in this way.  Then we can study what
happens to the quantum problem as we add operators corresponding
to higher frequency fluctuations in the
quantum fields back into the fully non-linear problem.

Another interesting direction is to pixelize a problem initially
quantized in a region extending over several horizon
volumes.  Since the pixelization to volumes smaller than the scale
set by the horizon during inflation certainly leads to coupling
terms between the pixels, there will be analogous (presumably
weaker) couplings between what will now be neighboring horizon-size
pixels.  The interesting question here is whether or not the
scale factors in neighboring volumes can get very far from one
another without causing observable changes in the CMB fluctuation
spectrum seen by an observer in any individual volume.  In other
words, can we---within the context of chaotic or
eternal inflation---put experimental limits on how near to us a
very different universe from ours can be, without
leaving a visible imprint on the CMB radiation in our
universe?  If such limits can be found, we will have found a way
to see the unseeable.

\section{Summary}

We have shown how to fully quantize the
theory of inflation and the computation of $\delta\rho/\rho$ by
working in a fixed gauge ({\it i.e.\/} co-moving coordinates) and making a
simple change of variables.  Our focus was on the formulation of
the part of the problem that involved the spatially constant
fields.  As we demonstrated, in both the classical and quantum theory,
working in a fixed gauge yields only two of the four relevant
Einstein equations as equations of motion. In the classical
theory we showed that the Friedmann equation and its time
derivative must be treated as constraints whose constancy in time
requires a proof.  This proof followed from the two equations of motion we
did have.  Next, we showed that, in the quantum version of the theory,
the same two Einstein equations appear as operator equations of
motion.  We then argue that the simplest constraint equation, which corresponds to the
Wheeler-DeWitt equation, cannot be used to define the space of
physical states.  We then show that there exist other possible choices for constraint
operators, but choosing them automatically makes the constraint equation different
from the Wheeler-DeWitt equation.  From this it follows that
identifying the Friedmann equation with the condition that the Hamiltonian
annihilates physical states is always incorrect.  In the paper which follows this one
we show explicitly how all of this plays out for the case of de-Sitter space.

\section{Acknowledgements}

MW would like to express his gratitude to J.~D.~Bjorken for dragging him,
kicking and screaming, into looking at this
fascinating problem.

\end{document}